\documentclass[preprint,12pt]{elsarticle}
\usepackage{amssymb}
\usepackage{amsmath}
\usepackage{graphicx}
\usepackage{tabularx, multirow}
\usepackage{booktabs}
\usepackage{url}
\journal{}

\begin{document}

\begin{frontmatter}



\title{Improving Performance in Colorectal Cancer Histology Decomposition using Deep and Ensemble Machine Learning}

\author[inst1]{Fabi Prezja}
\author[inst3,inst4]{Leevi Annala}
\author[inst1]{Sampsa Kiiskinen}
\author[inst1,inst5]{Suvi Lahtinen}
\author[inst1]{Timo Ojala}
\author[inst7,inst8]{Pekka Ruusuvuori}
\author[inst10,inst11]{Teijo Kuopio}

\affiliation[inst1]{organization={University of Jyväskylä, Faculty of Information Technology},
            city={Jyväskylä},
            postcode={40014}, 
            country={Finland}}

\affiliation[inst3]{organization={University of Helsinki, Faculty of Science, Department of Computer Science },
            city={Helsinki},
            country={Finland}}
            
\affiliation[inst4]{organization={University of Helsinki, Faculty of Agriculture and Forestry, Department of Food and Nutrition },
            city={Helsinki},
            country={Finland}}

\affiliation[inst5]{organization={University of Jyväskylä, Faculty of Mathematics and Science, Department of Biological and Environmental Science},
            city={Jyväskylä},
            postcode={40014}, 
            country={Finland}}

\affiliation[inst7]{organization={University of Turku, Institute of Biomedicine, Cancer Research Unit},
            city={Turku},
            postcode={20014}, 
            country={Finland}}

\affiliation[inst8]{organization={Turku University Hospital, FICAN West Cancer Centre},
            addressline={}, 
            city={Turku},
            postcode={20521}, 
            country={Finland}}

\affiliation[inst10]{organization={University of Jyväskylä, Department of Biological and Environmental Science},
            city={Jyväskylä},
            postcode={40014}, 
            country={Finland}}

\affiliation[inst11]{organization={Hospital Nova of Central Finland, Department of Pathology},
            city={Jyväskylä},
            postcode={40620}, 
            country={Finland}}

\begin{abstract}
In routine colorectal cancer management, histologic samples stained with hematoxylin and eosin are commonly used. Nonetheless, their potential for defining objective biomarkers for patient stratification and treatment selection is still being explored. The current gold standard relies on expensive and time-consuming genetic tests. However, recent research highlights the potential of convolutional neural networks (CNNs) to facilitate the extraction of clinically relevant biomarkers from these readily available images. These CNN-based biomarkers can predict patient outcomes comparably to golden standards, with the added advantages of speed, automation, and minimal cost. The predictive potential of CNN-based biomarkers fundamentally relies on the ability of CNNs to accurately classify diverse tissue types from whole slide microscope images. Consequently, enhancing the accuracy of tissue class decomposition is critical to amplifying the prognostic potential of imaging-based biomarkers. This study introduces a hybrid deep transfer learning and ensemble machine learning model that improves upon previous approaches, including a transformer and neural architecture search baseline for this task. We employed a pairing of the EfficientNetV2 architecture with a random forest classification head. Our model achieved 96.74\% accuracy (95\% CI: 96.3\%-97.1\%) on the external test set and 99.89\% on the internal test set. Recognizing the potential of these models in the task, we have made them publicly available.
\end{abstract}


\end{frontmatter}


\newcommand{\qvec}[1]{\textbf{\textit{#1}}}

\section{Introduction}
\label{sec:intro}
Cancer comprises diseases marked by rapid, uncontrolled growth of abnormal cells that form malignant tumors. These cells can detach, spread, and form new tumors in distant body parts, a process known as metastasis, which is the primary cause of cancer-related deaths \cite{qian2017cancer}. The World Health Organization reports that cancer is a leading global cause of death, responsible for one in six deaths \cite{WHO2022}. The most common areas for cancer to initially develop are the breast, lung, colon, and prostate.

Colorectal Cancer (CRC) is the third most common yet second deadliest cancer \cite{ColorectalCancerAlliance2022}. American Cancer Society data indicates 56\% of patients are diagnosed at stages where cancer has begun to metastasize \cite{malik2019colorectal,parveen2015cancer}. Early detection and treatment are paramount \cite{schiffman2015early}. Machine vision advancements, especially through deep neural networks \cite{lecun2015deep}, have improved automatic cancer and other disease classification \cite{kather2019predicting,bychkov2018deep,skrede2020deep,calimeri2017biomedical,frid2018gan,prezja2022deepfake,prezja2023adaptive}. Despite the technological progress, healthcare professionals still need to examine histologic samples to confirm diagnoses and assess tumor stages. Hematoxylin and Eosin (HE) staining is typically used to highlight key histopathological features in these tissues\cite{prezja2022h,khan2023effect}.

CRC patients are categorized into different groups to tailor their treatment and surveillance strategies. Grouping relies on multiple factors, including clinical outcomes, tumor genetics, quantitative biomarkers, clinical data, and histopathological and molecular analyses of the tumor. Many biomarkers stem from molecular and genetic tests \cite{kurland2012promise,spratlin2009clinical,o2008quantitative,waldman2009quantitative}. Recent insights into tumor immunology have revealed the critical role of the tumor microenvironment in tumor growth. Therefore, discovering novel predictive and prognostic biomarkers that effectively identify tumor characteristics is crucial.

In recent times, the first quantitative biomarkers based on deep learning have been extracted from HE stained whole-slide images \cite{skrede2020deep,kather2019predicting,danielsen2018prognostic,kather2019deep,sirinukunwattana2021image,echle2021deep}.A clinically relevant biomarker derived from Convolutional Neural Networks is a quantifiable indicator extracted from medical images (like histology slides) using deep learning techniques. This biomarker can independently predict clinical outcomes, such as survival rates or disease progression, and may provide insights into the biological processes of a disease. Kather and colleagues \cite{kather2019predicting} were the first to use deep learning to identify a biomarker for stages III and IV of CRC. This novel biomarker exhibited performance comparable to the existing gold standards for determining CRC outcomes \cite{sobin2011tnm,isella2015stromal} and could be automatically generated from images, saving time and resources. Conversely, known biomarkers (MSI, BRAF and KRAS) were also predicted \cite{wagner2023transformer} with deep learning transformers \cite{vaswani2017attention}. The approach greatly improved current methods for detecting microsatellite instability in surgical samples and achieved clinical-level accuracy in colorectal cancer biopsies, a significant finding in the field \cite{ruusuvuori2023deep}.

In their seminal research, Kather et al.\cite{kather2019predicting} applied convolutional neural networks (CNNs) \cite{lecun1995convolutional} to identify nine distinct tissue classes from HE-stained whole-slide images. Their methodology led to a noteworthy classification accuracy of 94.3\% on their external testing data. They then compiled output layer neuron activations into a single weighted score, named 'Deep Stroma', and tested this new CNN-biomarker for outcome prediction in new patient cohorts. They discovered that the 'Deep Stroma' score was a significant prognostic factor, especially in patients with advanced tumor stages (UICC IV). The CNN-biomarker was significantly prognostic in all tumor stages, while manual pathologist annotations and cancer-associated fibroblast (CAF) scores were not. Kather and his team prioritized the model with the highest accuracy, as it directly enhanced the quality and applicability of the new prognostic CNN-biomarker. 

Subsequent studies \cite{peng2019multi,qi2021identification,shen2022randstainna,wang2021accurate,yang2021medmnist,yang2021medmnist,shuai2022few,ghosh2021colorectal,schuhmacher2022framework} have made notable strides in improving accuracy. Nevertheless, some have either reported results not improving upon the original Kather, et al.\cite{kather2019predicting} model or faced challenges with incompatible output layer specifications and different validation methodologies. In our prior study \cite{prezja2023improved}, we refined the model design originally proposed by Kather et al.\cite{kather2019predicting}, achieving improved results to those previously reported. Building on that foundation, the current study introduces a new model that improves classification accuracy for this task and further builds upon previous solutions. Utilizing the EfficientNetV2 CNN architecture \cite{tan2021efficientnetv2} which has demonstrated notable capabilities in microscope image analysis\cite{raza2023lung,byeon2022automated,kallipolitis2021ensembling,munien2021classification} and in conjunction with the random forest ensemble algorithm \cite{breiman2001random}, we introduce a hybrid deep and ensemble model.

\section{Materials and Methods}
\label{sec:methods}
\begin{figure}[!ht]
\includegraphics[width=\textwidth]{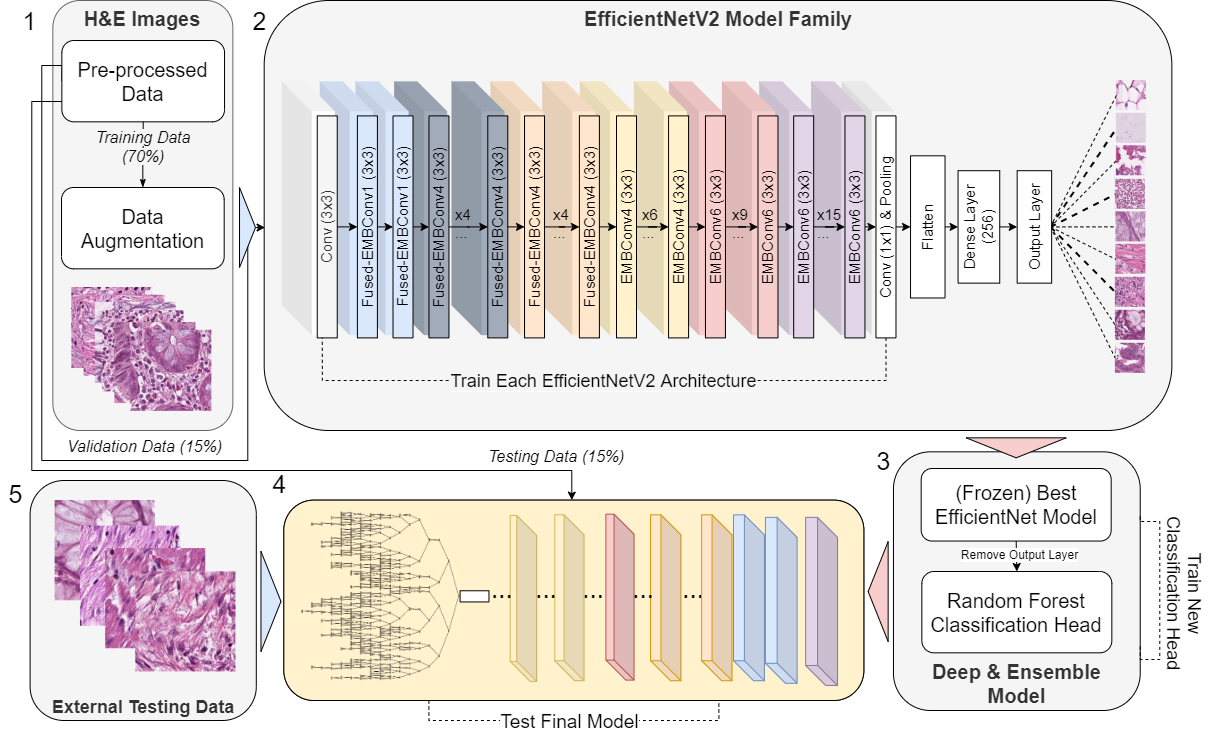}
\caption{Methodological Pipeline for Model Development. Blue arrows indicate data flow, red arrows represent model transfer, and numerical markers show the operational sequence.}
\label{flow}
\end{figure}

In this Methodology section, we detail our research process, outlining our data handling, image augmentation techniques, and the use of Convolutional Neural Networks. As shown in Figure \ref{flow} we focused on EfficientNetV2 for image classification, supplemented by the Random Forest method to generate the final deep ensemble model for refined predictions. These models were further evaluated against Transformer\cite{dosovitskiy2020image} and Auto-Keras baselines. The process began with the acquisition of HE-stained colorectal cancer data as the first step. In the second step, EfficientNetV2 models were trained for tissue classification. Once the best-performing EfficientNetV2 model was identified, it was frozen in the third step, and a new random forest classifier was trained using the learned features. The fourth step involved evaluating the hybrid model using both test and external datasets. Additionally, results from this model were compared with those obtained from Autokeras NAS and ViT transformers.

\subsection{Data Acquisition and Pre-processing}
\begin{figure} [!ht]
\includegraphics[width=\textwidth]{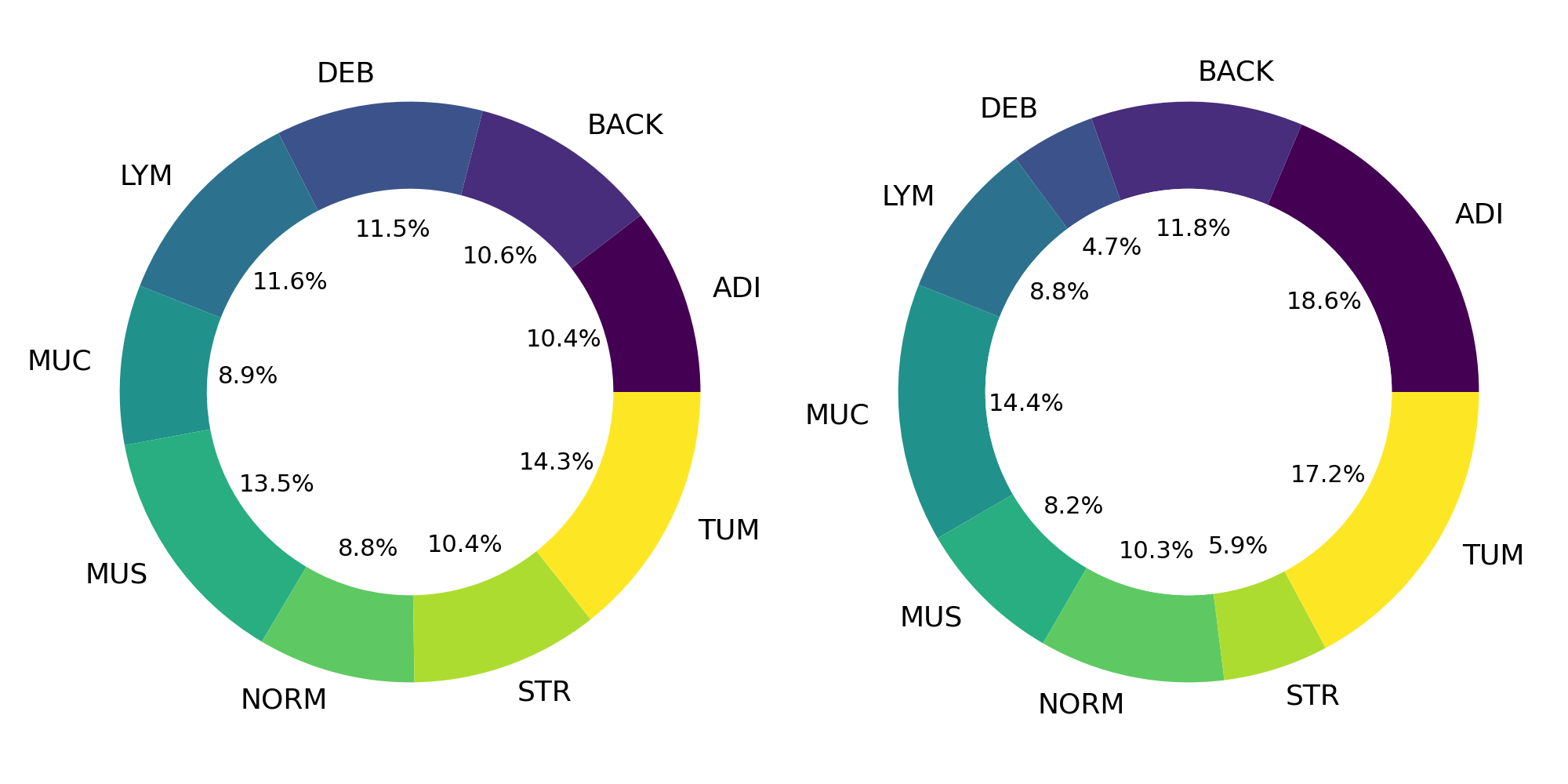}
\caption{Figure \ref{pie} illustrates the distribution of data from the source \cite{kather2019predicting}. On the left, the inner ring displays the percentage of data allocated for training, along with class labels on the outer ring. The right side shows a similar breakdown for the external-testing data.} \label{pie}
\end{figure}

\begin{figure} [!ht]
\includegraphics[width=\textwidth]{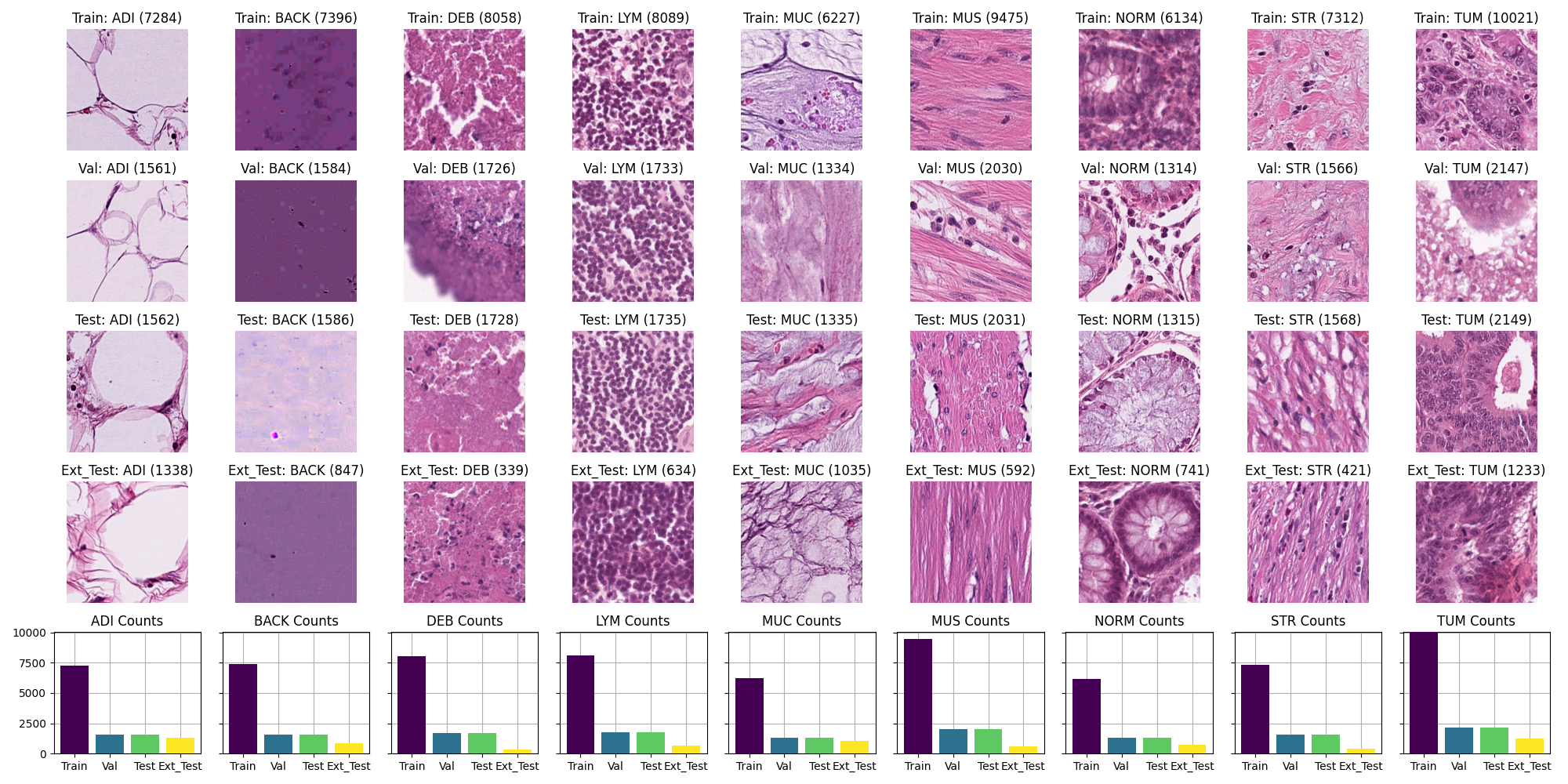}
\caption{Tissue sample tiles and distribution across dataset partitions for nine tissue types. Each row represents a different dataset partition (Training, Validation, Test, and External Test), denoted as Train, Val, Test, and Ext. Test, respectively. Below each tissue type column are bar graphs displaying the count of images in each dataset partition. } 
\label{tissues}
\end{figure}

The data utilized in this study were collected by the National Center for Tumor Diseases (NCT) in Heidelberg, Germany, and the University Medical Center Mannheim (UMM) in Mannheim, Germany. This comprehensive dataset has been made publicly available by Kather et al.\cite{kather2019predicting} and consists of 100,000 non-overlapping image tiles derived from 86 HE-stained tissue slides\cite{kather2018100}. The image tiles, each measuring 224 x 224 pixels, were normalized using the Macenko method\cite{macenko2009method}. These images span nine distinct classes: 1) adipose tissue (ADI); 2) background (BACK); 3) debris (DEB); 4) lymphocyte (LYM); 5) mucus (MUC); 6) smooth muscle (MUS); 7) normal colon mucosa (NORM); 8) cancer-associated stroma (STR); 9) CRC Epithelium (TUM). A detailed breakdown of the class distribution can be found in Figure \ref{pie}, and representative images for each class are displayed in Figure \ref{tissues}.

The dataset was partitioned into a training, validation, and testing set containing 69996, 14995, and 15009 images, respectively. This distribution corresponded to 70\% of the original data for training, 15\% for validation, and 15\% for testing. We also incorporated the external testing set from the original work by Kather et al.\cite{kather2019predicting}, consisting of 25 CRC HE slides from the NCT biobank, providing an additional 7180 image patches, code named (CRC-VAL-HE-7K)\cite{kather2018100}. Figure \ref{tissues} visualizes the number of images in each class for the training and external testing data.

\subsection{Data Augmentation}
Data augmentation, a technique to artificially enhance the diversity of training images \cite{shorten2019survey}, is applied by randomly transforming some images before they are fed into the training. Simple examples include random image rotation or shifting. When using multiple augmentations, the methods are combined to increase the possible variations. In this study's context, we employed six data augmentation methods bundled in the 'advanced' preset of the Deep Fast Vision Repository \cite{fabprezja_2023}. Each method and its specific configuration is detailed in Table \ref{tab:aug}. The same configuration was used for the training of the ViT transformer\cite{dosovitskiy2020image} from the ViT Keras\cite{morales_vit-keras} Library. Augmentation techniques were applied randomly to different images and in random combinations, in real-time as images were batched for training (with replacement).

\begin{table}[!ht]
\centering
\caption{A detailed description of the various image augmentation techniques utilized during the training of our neural networks.}
\label{tab:aug}
\scriptsize  
\begin{tabularx}{\textwidth}{lXX}
\toprule
\textbf{Augmentation Technique} & \textbf{Technique Explanation} & \textbf{Training Settings} \\
\midrule
Rotation & Rotates the image in the plane. & Allows image rotation up to a maximum of 40 degrees. \\
Width Shift & Shifts the image horizontally. & Allows a shift of up to +- 45 pixels along the width. \\
Height Shift & Shifts the image vertically. & Allows a shift of up to +- 45 pixels along the height. \\
Shear & Distorts the image by "stretching" it either horizontally or vertically, providing a 'skewed' perspective. & Shearing of the image is allowed up to a maximum angle of 0.2 degrees. \\
Zoom & Changes the image's apparent distance, making it seem closer or farther away. & Allows for a maximum zoom of up to 20\%. \\
Horizontal Flip & Creates a mirrored version of the image, providing a reflection along the vertical axis. & Only horizontal flipping is supported. \\
\bottomrule
\end{tabularx}
\end{table}

\subsection{Convolutional Neural Networks}
One of the bedrocks of the recent surge in deep learning is the Convolutional Neural Network (CNN) \cite{lecun1995convolutional}. A particular type of neural network, CNNs are often deployed for tasks in computer vision. They utilize the operation of convolution between an input and a filter-kernel. The kernels or filters are moved over the inputs to create feature maps, which represent highlighted features of the input. Different feature maps can be aggregated to form higher-level feature maps corresponding to more complex concepts. In a formal context \cite{GoodBengCour16}, given an image $\qvec{I}$ with dimensions $m\times n$ and a filter-kernel $\qvec{K}$ with dimensions $q\times r$, we generate the feature map $\qvec{F}$ via convolution along the axes $m, n$ with the kernel $\qvec{K}$ as follows:

\begin{equation}\label{eq:cnn}
\qvec{F}(m,n)=\sum_{q}\sum_{r}\qvec{I}(m,n)\qvec{K}(m-q,n-r)
\end{equation}

Typically, the values in the feature map undergo a transformation by an activation function. One common example is the rectified linear unit activation function \cite{nair2010rectified} (ReLu) which transforms all negative values to zero.

\subsection{EfficientNet Architecture}
EfficientNet \cite{tan2020efficientnet} is a model that has gained traction in machine learning applications in recent years. The architecture of EfficientNet has been designed to systematically scale all dimensions of depth, width, and resolution.

The underlying idea of EfficientNet is based on the compound scaling method. This strategy aims to maintain a balance between the network depth (how many layers it has), width (how wide are the layers), and resolution (input image size). A set of fixed scaling coefficients guides this balance. Formally, for a baseline EfficientNet-B0, if $\alpha, \beta, \gamma$ are constants which maintain the balance, and $\phi$ is the user-specified coefficient, we can scale depth $d$, width $w$, and resolution $r$ of the network according to:

\begin{equation}\label{eq:enet}
d = \alpha^{\phi} d_0, \quad w = \beta^{\phi} w_0, \quad r = \gamma^{\phi} r_0
\end{equation}

Here, $d_0, w_0, r_0$ are the depth, width, and resolution of the base model, respectively.

A core component of EfficientNet is the MBConv block, inspired by the MobileNetV2\cite{sandler2018mobilenetv2} architecture. This block applies a series of transformations: a $1\times1$ convolution (expansion), followed by a depth-wise convolution (represented by a depth-wise separable convolution kernel $\qvec{D}$), a Squeeze-and-Excitation (SE) operation \cite{hu2018squeeze}, and another $1\times1$ convolution (projection). For an input image $\qvec{I}$, the transformation of the MBConv block, $T_{MB}$, can be represented as:

\begin{equation}\label{eq:mbconv_se}
T_{MB}(\qvec{I}) = \qvec{K}_2 \ast SE(\qvec{D} \ast (\qvec{K}_1 \ast \qvec{I}))
\end{equation}

In this equation, $\ast$ represents the convolution operation, $\qvec{K}_1$ and $\qvec{K}_2$ are the $1\times1$ convolutional filters, $\qvec{D}$ represents the depth-wise convolutional filter,  and $SE(\cdot)$ denotes the Squeeze-and-Excitation operation.  Each convolution is followed by an activation function. This efficient use of computational resources within the MBConv block, especially through the depth-wise convolution and the SE block, contributes significantly to the excellent performance of EfficientNet.

EfficientNetV2 \cite{tan2021efficientnetv2} introduces several key changes to the original EfficientNet architecture. The depth, width, and resolution scaling remain the same as in the original EfficientNet. However, the use of the Fused-MBConv block, which combines the initial $1\times1$ convolution and the depth-wise convolution into a single $3\times3$ convolution operation, and then applies the Squeeze-and-Excitation (SE) operation followed by a final $1\times1$ convolution, is a significant modification.

The transformation of the Fused-MBConv block, $T_{FMB}$, can be expressed as:

\begin{equation}\label{eq:fmbconv}
T_{FMB}(\qvec{I}) = \qvec{K}_{2} \ast (SE(\qvec{K}_{f} \ast \qvec{I}))
\end{equation}

In this equation, $\qvec{K}_{f}$ is the $3\times3$ convolutional filter that combines the initial $1\times1$ convolution and the depth-wise convolution, $\mathbf{K}_{2}$ is the final $1\times1$ convolutional filter and $SE(\cdot)$ denotes the Squeeze-and-Excitation operation.Each convolution, followed by an activation, is part of a block that may contain a skip connection.

\begin{figure} [!ht]
\includegraphics[width=\textwidth]{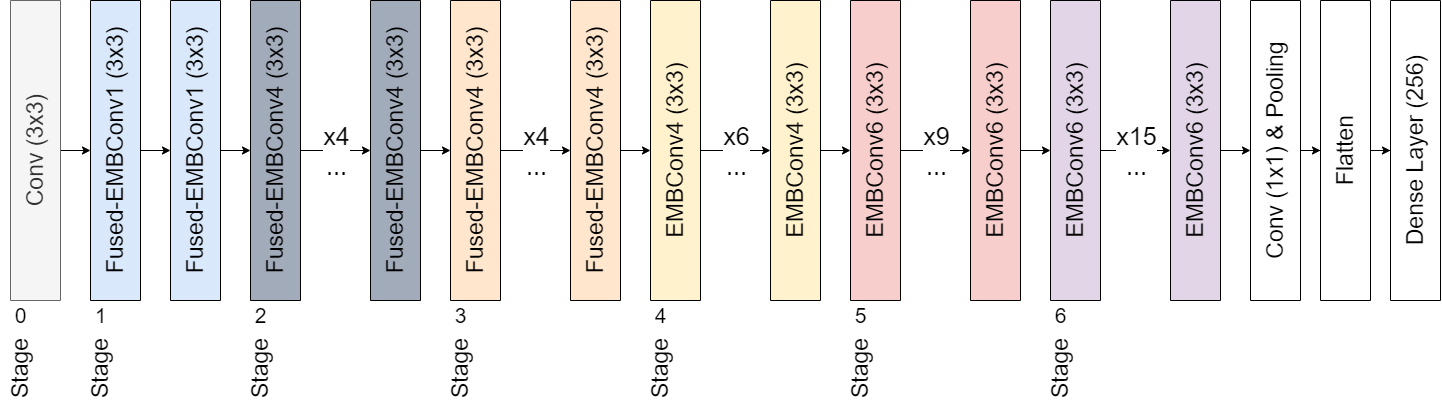}
\caption{Diagram illustrating the integrated architecture of the base EfficientNetV2 and our modifications after the pooling layer} \label{base}
\end{figure}

Another key change in EfficientNetV2 is the progressive learning method, which adaptively adjusts the regularization and image size during training. EfficientNetV2 models to train faster and achieve better parameter efficiency than the original EfficientNet, even outperforming transformer models on key vision tasks, such as Image-Net Classification. Figure \ref{base} illustrates the EfficientNetV2's base architecture (B0).

In our experiment, we extensively explored the EfficientNetV2 family of models, ranging from the relatively compact EfficientNetV2-B0 to the more complex EfficientNetV2-M. We supplemented the architecture by incorporating a flattening layer, followed by a densely connected layer of 256 neurons. This configuration emerged from an early-stage grid search on training data (split into 60\% train and 40\% validation), which tested various setups between 64 to 512 neurons (in steps of 64 neurons) and between 1 to 3 layers of depth. A single layer with 256 neurons yielded the best early validation accuracy. The grid search was conducted up to 22 epochs with early stopping, and we observed that results plateaued after 13 to 15 epochs (and early stopping activated). The search was conducted using EfficientNetV2B0, S, and M architectures to encompass models of low, mid, and high capacities. The dense layer units utilized an Exponential Linear Unit (ELU) as the activation function, set before the final output layer. For the training process, we employed Categorical Cross-Entropy Loss to optimize the model's performance, and trained for 15 epochs each model family for the main experiment. The minimum validation loss was used as the early stopping criterion. All EfficientNet model training was performed with the Deep Fast Vision repository. The optimizer used was Adam \cite{kingma2014adam}, with a learning rate of $2 \times 10^{-5}$  and a batch size of 32.

\subsection{Vision Transformers}

Vision Transformers\cite{dosovitskiy2020image} (ViT) utilize the transformer architecture to perform computer vision tasks by processing images as a sequence of fixed-size patches. An image $\qvec{I}$ of dimensions $H \times W \times C$ is segmented into patches $\qvec{P}_i$ of size $P \times P \times C$, which are then linearly transformed into embeddings $\qvec{E}_i$ using a projection matrix $\qvec{W}_p$:

\begin{equation}
\qvec{X}_0 = \qvec{W}_p \cdot \text{Flatten}(\qvec{P}i) + \qvec{E}_{\text{pos}}
\end{equation}

Here, $\qvec{E}_{\text{pos}}$ are positional embeddings added to retain spatial information. The ViT encoder applies multi-headed self-attention and position-wise feed-forward networks to these embeddings, where the attention for a single head is calculated as:

\begin{equation}
\text{Attention}(\qvec{Q}, \qvec{K}, \qvec{V}) = \text{softmax}\left(\frac{\qvec{Q} \qvec{K}^T}{\sqrt{d_k}}\right) \qvec{V}
\end{equation}

This formula consolidates the image processing and self-attention steps into a single representation, encapsulating the core functionalities of ViTs. The design of ViT reflects a shift towards utilizing global context and attention-based mechanisms to interpret images, contrasting with the localized receptive fields typically employed in convolutional neural networks. This method has demonstrated effectiveness when trained on large-scale datasets, showcasing the potential of transformers to generalize and manage complex visual tasks without intensive domain-specific modifications. In this study we employed the ViT B32 architecture, which is a specific configuration of the Vision Transformer (ViT). In the ViT B32 model, "B" signifies the Base model size, which involves 12 transformer blocks (12 attention heads), connected to 256 hidden layer units, while "32" indicates that each image is segmented into 32x32 pixel patches. The ViT was connected to the exact classification head specification as the efficient net models (256 hidden layer units) . The ViT trained maintaining the exact same batch size and augmentation approaches. Training lasted for 22 epochs, extending beyond the initial 15 to provide an overhead in effective capacity (training time). The Rectified Adam\cite{liu2019variance} optimizer was used, which may provide faster and improved convergence in terms of both time and accuracy.

\subsection{Random Forests}

Random Forests \cite{breiman2001random} (RF) is an ensemble machine learning method that leverages multiple decision trees to make predictions. Its robustness stems from combining a multitude of weak learners, the individual trees, to form a strong learner, the Random Forest. The most typical form of decision tree uses binary decisions based on feature thresholds to make predictions. The prediction of a single decision tree, $T_d$, on an input vector $\qvec{I}$, can be mathematically expressed as:

\begin{equation}\label{eq:tree}
T_d(\qvec{I}) = \sum_{i=1}^{n} v_i\delta(f_i(\qvec{I}) < t_i)
\end{equation}

In this equation, $n$ is the total number of nodes in the tree, $v_i$ is the predicted value for the $i$-th node, $f_i(\qvec{I})$ represents the $i$-th feature of the input vector $\qvec{I}$, $t_i$ is the threshold for the $i$-th node, and $\delta$ is the indicator function.

A Random Forest aggregates the predictions from $D$ decision trees to make a final prediction. For regression problems, the decision tree outputs are typically averaged, while for classification, they are aggregated through a majority vote (mode). Formally, the prediction of a Random Forest, $F_D$, on an input vector $\qvec{I}$, can be represented as:

\begin{equation}\label{eq:forest}
F_D(\qvec{I}) = \frac{1}{D} \sum_{d=1}^{D} T_d(\qvec{I})
\end{equation}

In this equation, $T_d(\qvec{I})$ is the output of the $d$-th decision tree for input vector $\qvec{I}$, and $D$ is the total number of decision trees in the forest. The Random Forest model's strength lies in its ability to reduce overfitting compared to a single decision tree by averaging the results over many trees. In our experiment, we employed 400 estimators and default parameters from the scikit-learn library \cite{scikit-learn}. Ultimately, the best-performing EfficientNet (Deep) model, determined by validation accuracy, had its classification head replaced by the Random Forest. After training and validating the RF, it was evaluated once on the test and external test sets.

\subsection{t-Distributed Stochastic Neighbor Embedding}

The t-Distributed Stochastic Neighbor Embedding (t-SNE)\cite{van2008visualizing} is a tool for visualizing high-dimensional data in a lower-dimensional space, often two or three dimensions. It is particularly effective for visualizing complex data structures, such as those produced by deep learning models, in a way that preserves the relationships and structures within the data.

In our experiment for visualizing the external test data from the trained model, we used two dimensions with an automatic learning rate on the dense layer before the output layer of the best EfficientNetV2 model. Additionally, we sampled 160 examples per class from the lower-dimensional space and rasterized\cite{genofx,rastf}the projected space. These two approaches can be seen in results Figure \ref{rasttsne}, while Figure \ref{rasttsneall} demonstrates the replacement of all data points with their corresponding whole-slide image patches.

\subsubsection{Neural Architecture Search (NAS)}
Neural Architecture Search (NAS) is an automated process aimed at discovering an approximately optimal neural network architecture for a specific task by systematically exploring a range of possible configurations. In the context of Autokeras NAS, NAS involves evaluating different combinations of network layers, their connections, and hyperparameters to identify the architecture that yields improved performance on a given dataset.

\subsubsection{Autokeras NAS Configuration}

Autokeras NAS\cite{JMLR:v24:20-1355} is an open-source Auto-ML library for deep learning, built on top of Keras and offers functionalities like automatic model selection and hyperparameter tuning. The library can handle various data types, including images, text, and structured data. For our experiment, we initiated a search incorporating instance normalization block, image-augmentation, ResNet V2 blocks pre-trained with ImageNet, all followed by flatten and dense layer blocks. We trained for 15 epochs and ran a maximum of 6 trials; early stopping (with restore best weights) was employed, and the search focused on validation loss.The NAS process ran for 52 hours with multiple parallel instances in a P100 GPU cluster.

\section {Results}

\subsection{Classification}

As presented in Table \ref{tab:efirfnet}, we evaluated the performance of a series of models, including several EfficientNet architectures (without testing), the final hybrid model, and an Autokeras NAS baseline. The table provides a comprehensive summary of the training, validation, testing accuracy scores, and parameter counts for each model. Notably, the EfficientNet models consistently showcased a rise in accuracy scores corresponding to an increase in model complexity, emphasizing the positive association between the number of parameters and model performance. Among the investigated models, EfficientNetV2M emerged with the highest validation accuracy. This deep model subsequently served as the basis for our hybrid model, complemented by a Random Forest (RF) classification head. Our hybrid model underwent further evaluations on internal and external test sets to ascertain its robustness and applicability. The model accuracy scores of 99.89\% on the internal test set and 96.74\% on the external test set. The 95\% confidence interval was obtain with 1000 iterations of boot-strapping\cite{efron1992bootstrap}. In comparison, the Autokeras NAS and Transformer baseline models did not match the accuracy levels exhibited by the hybrid model.Table \ref{tab:scores} offers a detailed comparative analysis, juxtaposing the performance metrics of our model with those from other studies. Notably, our approach achieved improved results in both internal and external testing datasets.

\begin{table}[!ht]
\caption{Comparison of accuracy and parameter count across different EfficientNet models, the Autokeras NAS and Transformer-ViTB32 baseline models. The 95\% CI refers to the 95\% confidence interval from bootstrapping}
\label{tab:efirfnet}
\scriptsize
\begin{tabularx}{\textwidth}{lXXXXX}
\toprule
\textbf{Model Name} & \textbf{Training Accuracy} & \textbf{Validation Accuracy} & \textbf{Testing Accuracy} & \textbf{External Testing Accuracy} & \textbf{Parameter Count (M)} \\
\midrule
EfficientNetV2B0 & 99.5\% & 99.57\% & - & - & 21.98 \\
EfficientNetV2B1 & 99.5\% & 99.73\% & - & - & 22.99 \\
EfficientNetV2B2 & 99.51\% & 99.73\% & - & - & 26.43 \\
EfficientNetV2B3 & 99.59\% & 99.75\% & - & - & 32.20 \\
EfficientNetV2S  & 99.68\% & 99.79\% & - & - & 36.39 \\
EfficientNetV2M  & 99.72\% & 99.8\% & - & - & 69.21 \\
EfficientNetV2M + RF & 100\% & 99.91\% & 99.89\% & 96.74\% (95\% CI:96.3\%-97.1\%) & 69.21 \\
Autokeras NAS Model (Baseline) & 100\% & 98.47\% & 98.52\% & 94.21\% & 26.77 \\
Transformer-ViTB32 (Baseline 2) & 99.45\% & 99.32\% & 99.32\% & 92.55\% & 87.65 \\
\bottomrule
\end{tabularx}
\end{table}

\begin{figure}[!ht]
\centering
\includegraphics[width=\textwidth]{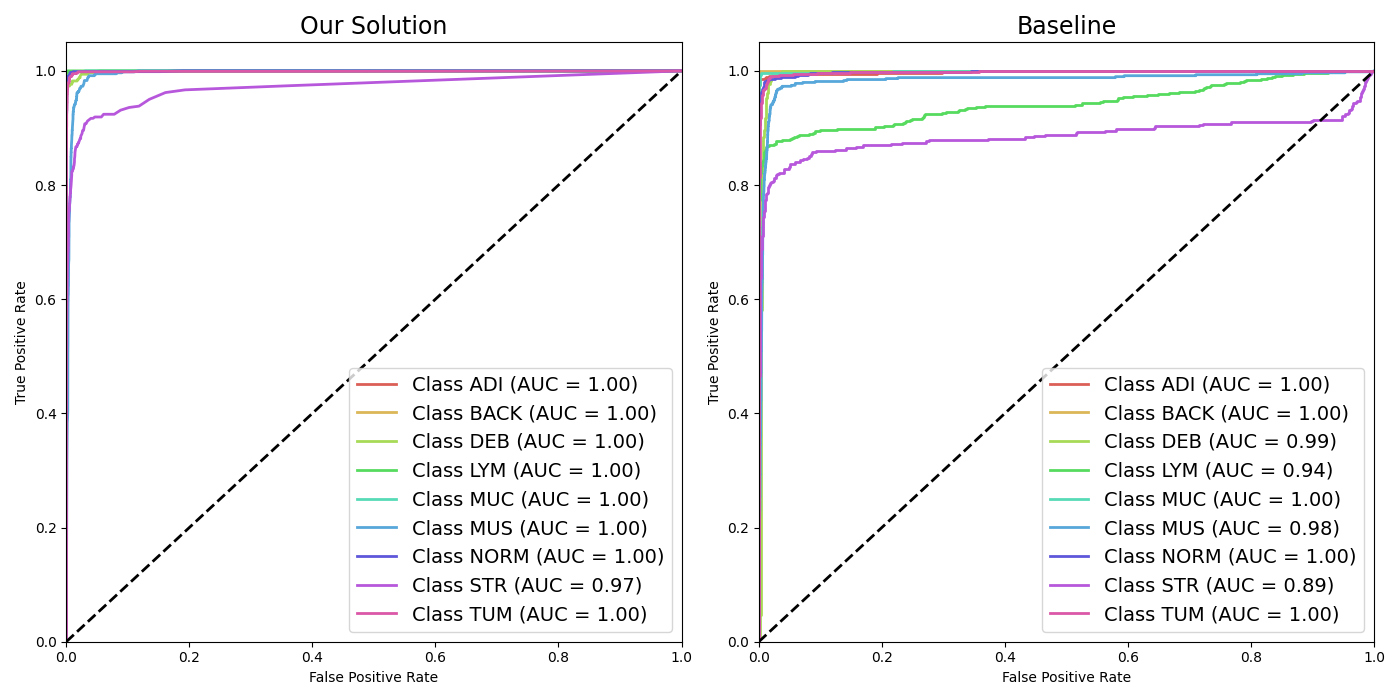}
\caption{Benchmarking EfficientNetV2M + RF hybrid vs. Autokeras NAS using one-vs-all ROC curves.} \label{roc}
\end{figure}

As shown in Figure \ref{roc}, using the one-vs-all scheme, our hybrid model and the Autokeras NAS baseline model's performance on the external testing set were assessed using the Area Under the Receiver Operating Characteristic (ROC) Curves. Our hybrid model displayed improved performance across all classes. It achieved the maximum AUC score of 1.00 for ADI, BACK, DEB, LYM, MUC, MUS, NORM, and TUM classes. For the STR class,  the AUC slightly dropped, and our model registered a score of 0.97. On the contrary, the Autokeras NAS baseline model, while demonstrating high AUC scores for ADI, BACK, MUC, NORM, and TUM, showed diminished performance for the remaining classes. The most notable dip was observed for class STR, registering an AUC of 0.89. Notably, the ROC curve for the Autokeras NAS baseline model intersected with the baseline after the 0.8 threshold. This intersection was not observed in the ROC curves of the hybrid model.

Table \ref{tab:scores} comprehensively compares accuracy scores across all relevant studies that employed the original training and external testing data. As shown, our model demonstrated improved performance, improving upon previous studies.

\begin{table}[!ht]
\centering
\caption{Comparative analysis of accuracy metrics and validation techniques across research studies.The 95\% CI refers to the 95\% confidence interval from bootstrapping}
\label{tab:scores}
\scriptsize  
\begin{tabularx}{\textwidth}{lXXXXX}
\toprule
\textbf{Research Work} & \textbf{Training Set (TR)} & \textbf{Validation Set (V)} & \textbf{Test Set (T)} & \textbf{Validation Approach} & \textbf{External Test Set (T2)} \\
\midrule
This study & \textbf{100\%} & \textbf{99.91\%} & \textbf{99.89\%} & TR-V-T-T2 & \textbf{96.74\% (95\% CI:96.3\%-97.1\%)} \\
Prezja, et al.\cite{prezja2023improved} & 99.4\% & 99.3\% & 99.5\% & TR-V-T-T2 & 95.6\% \\
Kather et al.\cite{kather2019predicting} & - & - & 98.7\% & TR-V-T-T2   TR-T2 & 94.3\% \\
Peng et al.\cite{peng2019multi} & - & - & - & TR-V-T-T2 & 95\% \\
Qi et al.\cite{qi2021identification} & - & 99\% & - & TR-V-T2 & 95\% \\
Shen et al.\cite{shen2022randstainna} & - & - & - & TR-V-T2 & 94.8\% \\
Wang et al.\cite{wang2021accurate} & - & - & - & TR-V-T2 & 94.8\% \\
Yang et al.\cite{yang2021medmnist} & - & - & - & TR-V-T2 & 91.1\% \\
Yang et al.\cite{yang2021medmnistb} & - & - & - & TR-V-T2 & 86.4\% \\
\bottomrule
\end{tabularx}
\end{table}

Furthermore, our study stands out as the only one providing a complete and transparent report of errors across all stages - from training through to external testing, and a confidence interval for the external testing benchmark. In evaluating prior studies, some \cite{tsai2021deep,shawesh2021enhancing} did not use conventional validation approaches, evident from the absence of validation and testing data to detect overfitting, including the search for parameters and hyperparameters on external testing data, which further complicate comparisons. Moreover, instances that employed techniques like few-shot learning and testing \cite{shuai2022few}, shuffling of external testing data within training data \cite{ghosh2021colorectal}, and using external testing as validation and testing with their own testing data \cite{schuchmacher2021framework},  also make it challenging to conduct a comparative analysis.

\begin{figure}[!ht]
 \centering
\includegraphics[width=\textwidth]{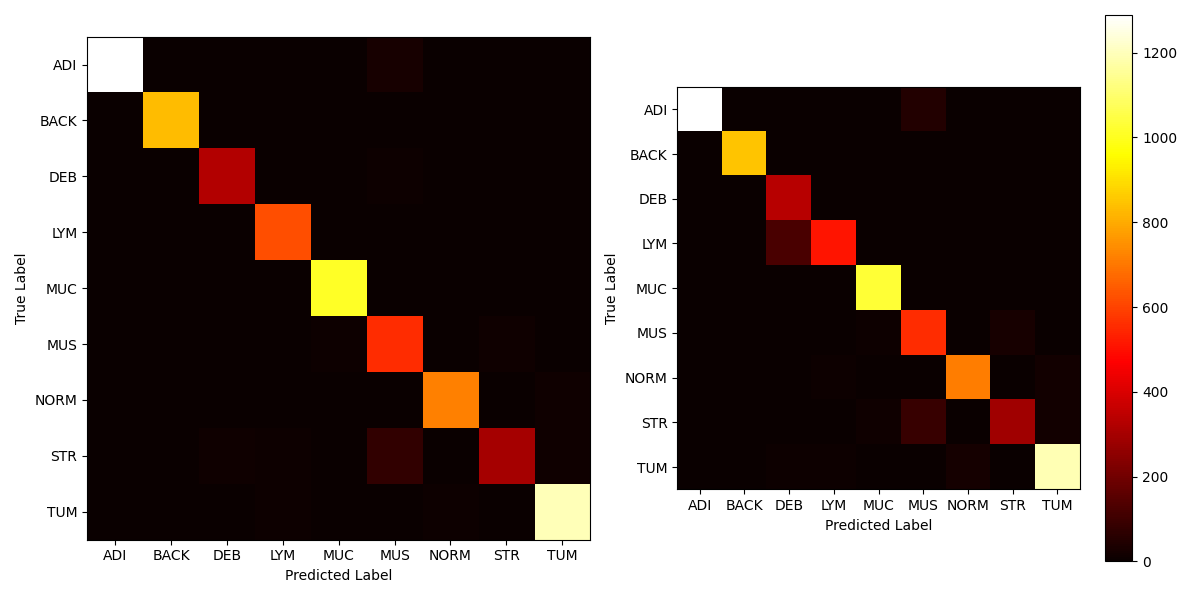}
\caption{Confusion matrices for external testing: Hybrid Model (left), Autokeras NAS baseline (right). Identical color format as Kather et al.\cite{kather2019predicting} for direct comparison.\label{confuse}}
\end{figure}

Figure \ref{confuse} demonstrates exceptional performance across most classes, with the vast majority of predictions falling on the diagonal, signifying correct classifications. Class BACK, LYM, and NORM are particularly noteworthy, which were nearly perfectly classified. However, the model exhibits some  confusion between classes MUC-MUS, and STR-MUS, suggesting shared features or characteristics that led to these misclassifications.

The confusion matrix from the original study conducted by Kather et al. revealed certain areas of confusion between particular classes, specifically LYM-DEB, MUS-MUC, NORM-TUM, TUM-STR, and STR-MUSC-MUC. When we compare this with the confusion matrix generated by our hybrid model, we see substantial improvements in classification accuracy, especially in the problematic areas identified in Kather's study. Figure \ref{confuse} (on the left) shows that our model successfully eliminated most of these previously reported confusions. For instance, our model has resolved the confusion between LYM-DEB, indicating its improved ability to distinguish between these two classes. Similarly, the confusion between MUS-MUC and NORM-TUM seen in Kather's study has also been resolved. One of the most substantial improvements was observed in the classification of LYM, where our model demonstrated almost perfect classification accuracy. In figure \ref{confuse} (on the right), the Autokeras NAS model, still achieving overall solid performance, underperformed in several areas relative to the hybrid model. Specifically, there is apparent confusion between classes DEB-LYM, and STR-MUS. It is also worth noting that, unlike the hybrid model, the Autokeras NAS model made several misclassifications between classes TUM and NORM.

The hybrid model's improved performance is most noticeable when dealing with classes ADI, MUS, LYM, NORM, and STR, suggesting that its architecture may be more adept at recognizing the nuanced features that distinguish these classes. These features are relevant to the Deep Stroma score calculation proposed by Kather.

\begin{figure}[!ht]
 \centering
\includegraphics[width=\textwidth]{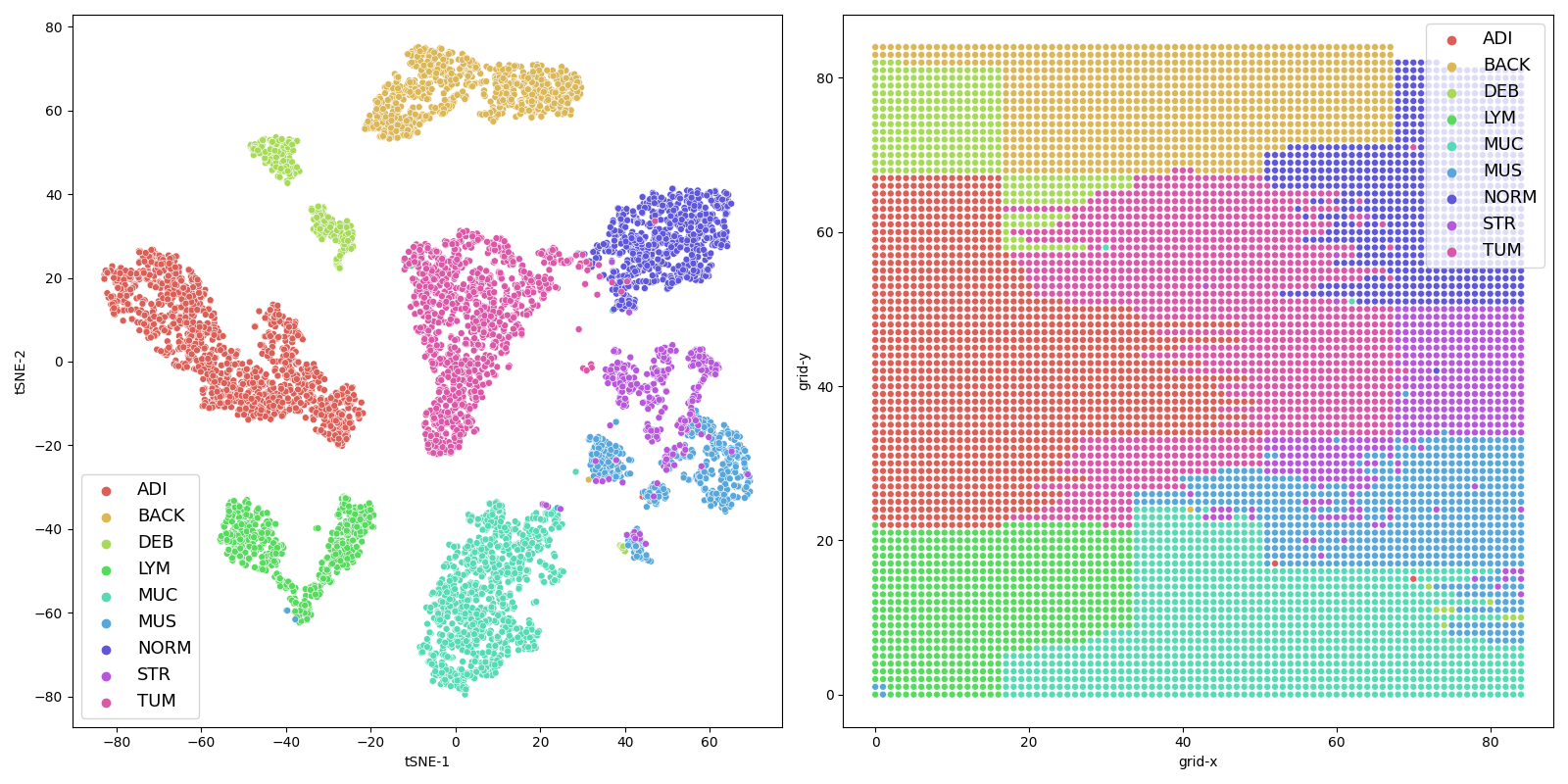}
\caption{Left: t-SNE projection illustrating near-optimal class separation with TUM class at center. Right: Rasterized t-SNE map preserving relative inter-class distances and eliminating inter-point distance.} \label{tsne}
\end{figure}

\begin{figure}[!ht]
 \centering
\includegraphics[width=\textwidth ]{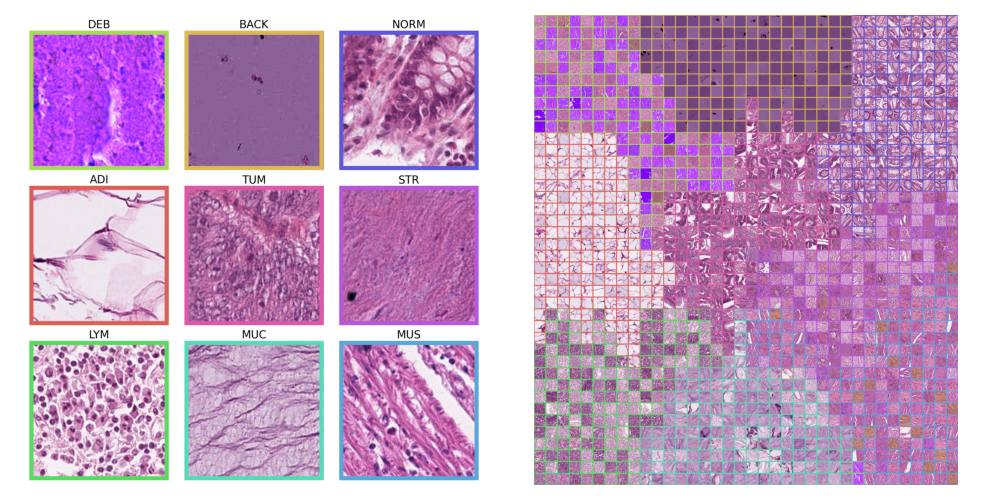}
\caption{Left:Nearest intra-class image to class mean in the t-SNE projection. Right: Rasterized t-SNE space with 160 samples per class, data points replaced with images. Boundary colors indicate class origin.} \label{rasttsne}
\end{figure}

\begin{figure}[!ht]
 \centering
\includegraphics[width=\textwidth]{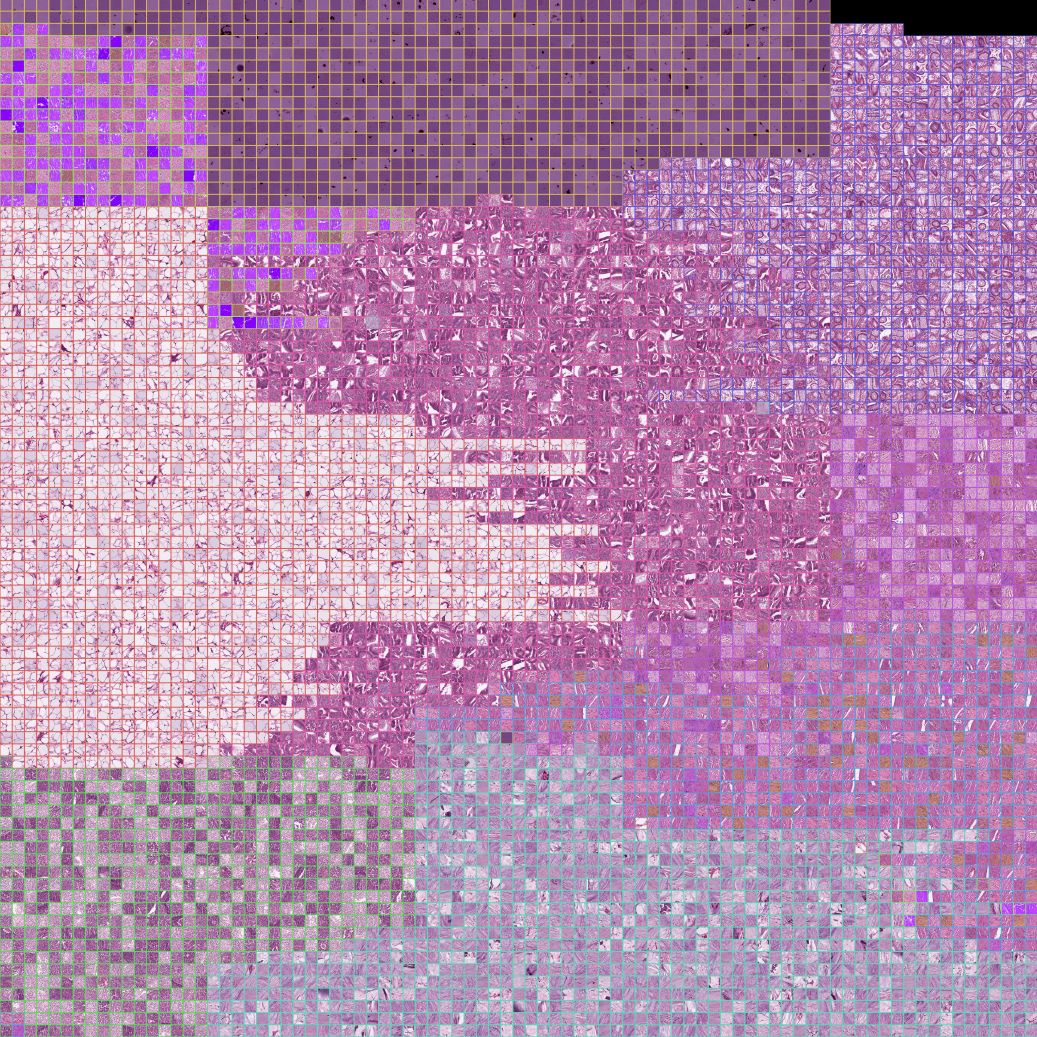}
\caption{Rasterized t-SNE visualization of all external test data showing class distributions. Black indicates blank grid space, boundary colors denote class origin. The full resolution is made available under data availability.} \label{rasttsneall}
\end{figure}

In the t-SNE visualization of the high-dimensional feature space (Figure \ref{tsne}), the classes align almost seamlessly around the tumor (TUM) class at the approximate center. This near-optimal separation of classes demonstrates the effectiveness of the hybrid model in distinguishing between various histopathological types. Notably, there is no observed fragmentation within the classes, underlining the consistency of the learned features even as a two-dimensional projection.
Moreover, the relative positioning of the classes aligns with histopathological anticipations. The TUM and normal (NORM) tissues, displaying histological similarities, are situated adjacently on the t-SNE map. Analogously, the proximity of the stroma (STR) and muscle (MUS) classes suggests vector space congruence. Transitioning the t-SNE map into a rasterized depiction preserves relative inter-class distances and condenses the visualization by eliminating inter-point space, enabling a more compact visualization of class distributions. Figure \ref{rasttsne} illustrates a single instance from each class closest to each class mean as projected by t-SNE. Concurrently, we subsample the t-SNE space with 160 samples per class to build a rasterized representation, substituting data-points with corresponding images. Figure \ref{rasttsneall} follows the same pattern but without sub-sampling, using all external test data.

Overall, these results demonstrate the robustness of the hybrid model's feature extraction and highlight its potential to provide clinically relevant insights.

\clearpage

\section {Discussion}
The present study was attempted to advance the decomposition classification of colorectal cancer by developing a novel hybrid model. Our methodology integrated EfficientNetV2M and Random Forest (RF) techniques, with the results demonstrating the model's improved performance over previously proposed solutions (Table \ref{tab:scores}). This was achieved by combining the strengths of both deep learning and traditional ensemble machine learning. In addition to achieving high accuracy rates, our hybrid model demonstrated improved performance across all classes when assessed using the Area Under the Receiver Operating Characteristic (ROC) Curves in a one-vs-all scheme (Figure \ref{roc}). The hybrid model achieved the maximum AUC score of 1.00 for classes ADI, BACK, DEB, LYM, MUC, MUS, NORM, and TUM. While the model performance slightly dropped for the STR class, it still registered an AUC of 0.97. The confusion matrices demonstrated the hybrid model's improved ability to correctly classify various histopathological types, even those that presented difficulties in previous studies, such as the original study by Kather et al\cite{kather2019predicting}.

Combining Random Forests with Convolutional Neural Networks (CNNs) represents a relatively unexplored area within medical imaging. Prior to this study, this approach was applied using a two-layer CNN trained from scratch with a binary target, specifically with MRI images \cite{khozeimeh2022rf}. Although promising, these initial applications did not fully explore the potential efficacy of combining more complex neural network architectures and leveraging transfer learning. Our study extended this methodology into a new domain by employing advanced CNN architectures such as EfficientNetV2 and incorporating transfer learning techniques along with a multi-class target. The results demonstrated improved task performance with these new parameters and more complex neural networks. This underscored the potential of combining traditional machine learning algorithms with state-of-the-art deep learning models for enhanced predictive accuracy in histopathological image analysis.

Progress in the task of colorectal cancer histology decomposition appears necessarily incremental, given that less than 5\% improvement remains to reach a maximum score in available testing data. However, this seemingly modest threshold is crucial due to the scalability potential of the system. Small enhancements in performance may lead to substantial benefits when applied across large datasets, where tens of thousands of histology slides are analyzed. This phenomenon mirrors trends observed in the ImageNet Challenge, where similar incremental improvements still drive researchers to choose newer architectures over older ones due to the competitive edge they provide. Likewise, in our study, we chose the EfficientNetV2 architecture for its proven incremental improvements over previous models in the ImageNet Challenge.

One significant limitation encountered in our study is the inability to apply Gradient-weighted Class Activation Mapping (Grad-CAM) \cite{selvaraju2017grad} for visualizing class activation regions in input images. Grad-CAM relies on the gradients flowing through the final convolutional layers to produce heat maps that highlight the areas of the image most relevant to the classification decision. However, when employing a Random Forest (RF) as the final classification head, the gradients necessary for generating these heat maps are not computable. Consequently, the integration of RF with CNNs, while effective in enhancing classification performance, inherently precluded the use of gradient-based visualization techniques such as Grad-CAM. This limitation suggests the need for alternative methods to visualize class-related activations in hybrid models that combi    ne deep learning and traditional machine learning algorithms.

The t-Distributed Stochastic Neighbor Embedding (t-SNE) visualization of the high-dimensional feature space demonstrated the effectiveness of the hybrid model in distinguishing between various histopathological types. Additionally, t-SNE revealed the relationships among class instances in the projected vector space, though the two dimensions of this vector space do not yet have a higher-level interpretation. To this end, we are providing the rasterized t-SNE plot in high definition to assist future medical experts in identifying the projected features. This effort aims to enhance the interpretability of the projected space, enabling medical professionals to gain better insights into the relationships and structures within the data.

In terms of data augmentation and validation, our findings suggest that robust data augmentation is a powerful tool for mitigating overfitting when paired with an appropriate validation technique. This conclusion is consistent with established practices, as data augmentation is a commonly used method in numerous classification studies. We believe that more advanced forms of augmentation, particularly those that utilize synthetic data from Generative Neural Networks \cite{mauricio2018high,jose2021generative,prezja2022deepfake,prezja2023synthesizing}, could also prove beneficial and may warrant a dedicated future study. As illustrated in Table \ref{tab:scores}, there is considerable variation in validation methods, which poses a challenge when validation, testing, and external testing datasets are missing. The practice of validation in this context could be significantly improved by adopting a standardized approach. From a limitation perspective, the training of such systems necessitates a substantial volume of annotated data. For potential advancements in these systems, the annotation of even more data by medical professionals may be required. Notably, several studies \cite{shen2022randstainna,yang2021medmnist,shawesh2021enhancing} are still pending peer review.

In reflecting upon the prevailing CRC scientific literature, we have noted considerable advancements achieved over a comparably short time. However, this progress has not been devoid of limitations. Firstly, our survey identified a distinct lack of uniformity in the evaluation methodologies employed across different systems. Emphasizing consistency and adherence to best practices during evaluation is crucial for reducing biases and enabling direct system comparisons. Moreover, estimates for metrics such as label noise are absent. Ascertaining estimates for label noise could be pivotal in establishing a benchmark for future comparative analyses.

In our study, the hybrid model’s deep section was trained using our new Deep Fast Vision open-source library \cite{fabprezja_2023}, whereas the baseline employed Autokeras NAS \cite{JMLR:v24:20-1355}. Both these libraries uniquely leverage AutoML in vision-related contexts with transfer learning capabilities. The hybrid model demonstrated improvement in scores compared to Autokeras NAS. Although Autokeras NAS employs ImageNet pre-trained blocks, such blocks function independently, and only low-level features are loaded. On the contrary, Deep Fast Vision utilizes the entire hierarchy of learned features from already validated architectures. This methodology gives Deep Fast Vision an expected early advantage over Autokeras NAS, allowing it to harness the transfer learning capability of previously validated deep learning models.

The combination of EfficientNet and Random Forests introduces a novel approach by integrating a cutting-edge deep learning architecture with a traditional machine learning algorithm, a specific pairing that has not been explored previously. In addition, to the best of our knowledge, this is the first instance where the EfficientNetV2 architecture is utilized in colorectal cancer (CRC) histopathology imaging. Intriguingly, the hybrid model outperformed a vision transformer baseline, which possessed notably greater parameter counts. These findings are both unique and promising. However, it is important to note a significant limitation: we were unable to utilize transformers as the classification head. The primary reason is that the input features to tabular transformers\cite{huang2020tabtransformer} would neither be categorical nor a mix of categorical and numeric. As a result, these features bypass the advantages of transformers' attention mechanisms and proceed directly to the MLP phase. Additionally, it is crucial to highlight that Random Forest cannot receive gradients directly from the CNN or its flattening layer, necessitating the application of the Random Forests asynchronously.

On Transformer models and Autokeras NAS baselines, it is essential to address the variations in performance between different architectural approaches and training strategies within this task. Specifically, the difference in performance between Transformers and Autokeras NAS was not surprising when considering the depth and scope of the neural architecture search (NAS) implemented by Autokeras. This approach extended the search duration significantly, optimizing for the most effective model configurations over a prolonged period. The NAS process ran for 52 hours with multiple parallel instances in a P100 GPU cluster. In contrast, EfficientNet, which was included in our hybrid model, had previously demonstrated improved performance over Vision Transformer (ViT) models in benchmarks like ImageNet. This result could explain the increased performance of EfficientNet over ViTs.  The performance of larger Transformer models remains untested in this context, suggesting that this topic warrants further investigation through a future study. Additionally, utilizing a transformer as a second baseline is crucial to ground and contrast our results, moving away from solely relying on automated machine learning (AutoML) solutions. Moreover, AutoKeras NAS was not constrained by a capacity threshold, and trained long enough to also gain more effective capacity (training time). This implies that the Autokeras NAS was capable of searching long enough to potentialy outperform both Transformer and EfficientNet models but did not surpass EfficientNet, indicating that factors other than model capacity were at play, which could warrant further investigation in future work.

In our study, we utilized Autokeras NAS for automated neural architecture search (NAS), specifically incorporating the ResNet block due to its availability of pre-trained weights from ImageNet. This decision was critical for maintaining a fair comparison across different models. ResNet blocks, being pre-trained, allow for the use of pre-learned features, which significantly accelerate the convergence process during the training phase. This is similar to the advantage that EfficientNet gains from its pre-training. Without utilizing blocks that come with pre-trained weights, comparisons with EfficientNet would naturally be skewed. EfficientNet benefits from a pre-training phase that enhances its initial performance and learning speed, attributes that would be lacking in a NAS-generated model built from scratch. By using the ResNet block within Autokeras NAS, we aimed to level the playing field, ensuring that all models being compared had similar advantages in terms of initial feature learning and convergence speed.

Concerning the quality of microscope images, factors such as the slide's quality and the presence of artifacts or pixel noise could significantly contribute to misclassifications. Normalization and contrast enhancement processes performed on tiles before their entry into the classifier might accentuate pixel noise or other non-tissue artifacts, such as dust or hair, thereby potentially skewing results. The 'Picasso' effect \cite{gliozzi2022combining} may compound these distortions in Convolutional Neural Networks (CNNs). Furthermore, instances, where a tile contains minimal or no relevant tissue and isn't labeled as background could also instigate misclassifications. Incorporating a background class can somewhat mitigate these issues, but this effect is only partial, and similar errors would be anticipated. To address this more comprehensively, employing adversarial augmentations\cite{prezja2023exploring} could help uncover additional vulnerabilities. Future systems could benefit from introducing pixel noise during the augmentation phase, ideally in a randomized manner and with replacement. Randomization and replacement are critical to prevent the introduction of biases and potential overfitting by the classifier. Lastly, the focus factor ('blur') can also influence results, particularly when it coexists with pixel noise. As a preventative measure, further augmentation involving a variety of blur intensities might be beneficial. Such recommendations are particularly crucial given the variability in focus and quality across different patient slides. Additionally, it is important to note that staining normalization\cite{clarke2017colour,vahadane2016structure,roy2018study,khan2014nonlinear,salehi2020pix2pix,prezja2022h} could also affect results and might warrant an independent future study to thoroughly assess its impact.

In our previous study \cite{prezja2023improved}, we highlighted the need for a deeper analysis of classifier probability profiles \cite{guo2017calibration}. While Kather's approach \cite{kather2019predicting} pinpointed one classifier, it did not address the influence of model probability calibration on the deep stroma score. Given that different architectures can produce varying probability profiles, even with similar performance metrics, there's a gap in understanding these profiles' effects on deep stroma and outcome prediction. Addressing this could open new avenues for research and refine our approach to model architectures and outcome predictions. Before advancing to bio-marker extraction and patient outcome assessment, it's imperative that these profiles undergo further validation.

In Kather, et al.\cite{kather2019predicting} from which we sourced our data, there was a stringent manual review process for all slides. Slides with pronounced artifacts, be it tissue folds or tears, were set aside. This exclusion underscores an inherent gap: our models, along with others, have not been trained with these challenges. In many real-world scenarios where such artifacts may exist, the desired performance might decline, emphasizing the importance of training datasets that mirror extensive real-world challenges. While remedies like affine augmentations might provide some mitigation, they fall short of fully addressing these challenges. Additionally, it should be noted that to our knowledge no further annotated data is available for additional training, refinement or validation.

\section {Conclusion}

Our study successfully developed a hybrid model that combines the EfficientNetV2 and Random Forest algorithms, notably advancing the task scores of automatic colorectal cancer tissue decomposition. Our model demonstrated high average accuracy and excelled in performance across all classes, as indicated by the AUC scores. It improved upon the results of previous research in this domain, including the seminal work by Kather et al.\cite{kather2019predicting}. 

It is imperative that this work is subjected to rigorous clinical validation before any deployment into routine clinical practice. Our research presents considerable potential in the classification of CRC slides and the prospect of improving CNN-based biomarkers that rely on classification accuracy. This improvement, in turn, could lead to improved predictions for CRC patient outcomes. 

Lastly, we have made a commitment to transparency and replicability by providing unrestricted access to our models.

\section {Data Availability}

Materials and best models from the current study are accessible in the Google Drive repository: \path{https://drive.google.com/drive/folders/}\\
\path{1ypFyU2V6ifRkLB6hRKlqb6C2D_fvL_5Y?usp=sharing}


 \bibliographystyle{elsarticle-num} 
 \bibliography{main}

\section*{Acknowledgements}
The authors extend their gratitude to Kimmo Riihiaho, Rodion Enkel and Leevi Lind.

\section*{Author contributions statement}
Conceptualization: F. P. \& T.K.; 
Methodology: F. P.; 
Investigation: F. P. \& T.K.; 
Data Curation: All authors; 
Formal analysis: All authors; 
Writing – original draft: F. P.; 
Writing – review \& editing: All authors.

\section*{Additional information}
 \textbf{Competing interests}
 All authors declare that they have no conflicts of interest.





\end{document}